\documentclass[12pt]{article}

\renewcommand{\Im}{\textrm{Im}\,}
\def\i{\mathrm{i}}
\def\e{\mathrm{e}}
\def\arg{\mathrm{Arg}\,}

\usepackage[dvips]{graphicx}
\usepackage{amsfonts,amsmath,amssymb}
\usepackage{times}

\begin{document}
      \title{Group delay in Bragg grating with linear chirp}
      \author{O.V. Belai, E.V. Podivilov, D. A. Shapiro\\
      Institute of Automation and Electrometry,\\
      Russian Academy of Sciences, Siberian Branch,\\
      Novosibirsk, 630090 Russia}

\maketitle

\begin{abstract}
An analytic solution for Bragg grating with linear chirp in
the form of confluent hypergeometric functions is analyzed
in the asymptotic limit of long grating. Simple formulas
for reflection coefficient and group delay are derived. The
simplification makes it possible to analyze irregularities
of the curves and suggest the ways of their suppression. It
is shown that the increase in chirp at fixed other
parameters decreases the oscillations in the group delay,
but gains the oscillations in the reflection spectrum. The
conclusions are in agreement with numerical calculations.

\noindent PACS {42.81.Wg; 78.66.-w}
\end{abstract}

\section{INTRODUCTION}
\label{sect:intro}

Optical filters based on fiber gratings attract particular
interest because of their applications in high-speed
lightware communications \cite{ptTAPC00}, fiber lasers
\cite{D01} and sensors \cite{FOS91}.
The Bragg reflector is
based on periodic modulation of the refractive index along
the line \cite{OK99,RK99}.
Gratings that have a nonuniform period along their length
are known as chirped. The theory of linearly chirped
grating holds the central place in the fiber optics.
Chirped grating is of importance because of its
applications as a dispersion-correcting or compensating
devices \cite{olO87}. The study of linearly chirped grating
is also helpful for approximate solution of more general
problem of complex Gaussian modulation \cite{ocSL04}. The
group delay as a function of wavelength is a linear
function with additional oscillations. For applications the
problem is to minimize the amplitude of regular
oscillations and the ripple resulting from the errors of
manufacturing \cite{ofcrSF05}.

The purpose of this work is to present and study a solution
of the equations for amplitudes of coupled waves in
quasi-sinusoidal grating with quadratic phase modulation.
The solution of coupled-wave equations is derived in terms
of the confluent hypergeometric functions. Their asymptotic
expansion in terms of Euler $\Gamma$-functions makes it
possible to obtain relatively simple formulas for
reflectivity and group delay. The simplification enables
analysis of irregularities of the curves and suggestions on
the ways of their suppression.

The paper is organized as follows. The equations for
amplitudes in the grating with quasi-sinusoidal modulated
refractive index are derived in Sec.~\ref{sect:2}. Their
analytic solutions are obtained and compared with numerical
results in Sec.~\ref{sect:3}. The asymptotic behavior is
treated in Sec.~\ref{sect:4}. Some estimations and
qualitative explanations are presented in
Sec.~\ref{sect:5}. Possible methods to suppress the
oscillations are summarized in Sec.~\ref{sect:6}.

\section{Equations for slow amplitudes}\label{sect:2}

Consider a single-mode fiber with  the weakly modulated refractive index $n+\delta
n(z)$. Steady-state electric field $E(z)$ satisfies
one-dimensional Helmholtz equation
\begin{equation}\label{Helmholtz}
\frac{d^2E}{dz^2}+k^2\left[1+\frac{2\delta n(z)}{n}
+\left(\frac{\delta n(z)}{n}\right)^2\right]E=0,
\quad k=\frac{\omega n}{c},
\end{equation}
where $z$ is the coordinate, $k$ is the wavenumber in glass
outside the grating, where $\delta n(z)=0$, $\omega,c$ are
the frequency and speed of light. The addition to mean
refractive index may be a function with phase and amplitude modulation.
A family of analytical solutions for amplitude
modulation was obtained in \cite{ocS03}.
Below we treat a case of phase modulation
\begin{equation}\label{quasi-sinusoidal}
\frac{\delta n(z)}{n}=2\beta\cos\theta(z),
\end{equation}
where $\theta(z)$ is the phase, constant $\beta$ is the
modulation depth.  Since $\beta\ll1$ we
neglect the quadratic term in (\ref{Helmholtz}). The phase
is general quadratic function
\begin{equation}\label{quadratic}
    \theta(z)=\alpha z^2/2+\kappa z +\theta_0,
\end{equation}
where $\kappa$ is the frequency of spatial modulation at
$z=0$, $\theta_0$ is the constant phase shift. The
condition of slow phase variation is
\begin{equation}\label{slow-variation}
    \left\vert\frac{d\theta}{dz}-\kappa\right\vert\ll\kappa.
\end{equation}

\begin{figure}\centering
\includegraphics[width=0.7\textwidth]{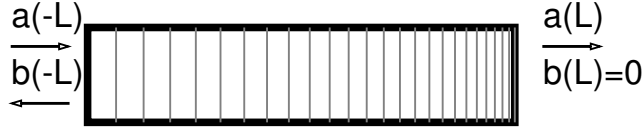}
\caption{The statement of the scattering problem.}
\label{fig:envelope}
\end{figure}

Let us introduce complex amplitudes $a,b$ of waves running in positive
and negative directions
\[
E=a\e^{\i kz}+b\e^{-\i kz}.
\]
Keeping only resonant terms and
neglecting the parametric resonance of higher orders at the detuning
\begin{equation}\label{resonance-condition}
q=k-\kappa/2\ll k_0=\kappa/2,
\end{equation}
we get the equations for coupled waves
\begin{equation}\label{interaction}
a'=\i k_0\beta \e^{-2\i k z+\i\theta(z)}b,\quad
b'=-\i k_0\beta\e^{2\i k z-\i\theta(z)}a,
\end{equation}
where prime denotes $z$-derivative.

Set (\ref{interaction}) conserves $|a|^2-|b|^2$, since the
signs in right-hand sides of equations are different. The
same equations with identical signs conserve the sum of
populations $|a|^2+|b|^2$ and describe the amplitudes of
probability in two-state quantum system. The exact
solutions in this case are of importance in quantum optics,
then they are studied in details \cite{AE87,praCM00}.
Within the limits of resonance approximation
(\ref{resonance-condition}) we replace $k$ in front of
exponents (\ref{interaction}) by $k_0$.

Finding the derivatives of (\ref{interaction}) with respect
to $z$ we get complex conjugated second-order equations
\begin{equation}\label{second-order}
    a''-\i\alpha\left(z-z_0\right)a'-k_0^2\beta^2 a=0,\quad
    b''+\i\alpha\left(z-z_0\right)b'-k_0^2\beta^2 b=0.
\end{equation}
Here $z_0=(2k-\kappa)/\alpha=2q/\alpha$ is the coordinate
of resonance point for the wave with wavenumber $k$. It is
the turning point where the wave with given $q$ is
reflected. The parametric resonance for central wavenumber
$q=k-\kappa/2=0$ occurs at $z=0$. Let $\alpha>0$, then for
the red detuning $q<0$ we have $z_0<0$, in opposite case
$q>0$ of blue detuning $z_0>0$.

Consider the Bragg grating written in the interval $-L\leqslant z\leqslant L$.
The problem of left reflection coefficient calculation is
illustrated by Fig.~\ref{fig:envelope}. Boundary conditions are defined by the
scattering problem statement.  We set amplitude $b$ at the
right end equal to zero
\begin{equation}\label{boundary}
    b(L)=0
\end{equation}
 and get the reflection and transmission coefficients
\begin{equation}\label{transmission}
    r=\frac{b(-L)}{a(-L)},\quad t=\frac{a(L)}{a(-L)}.
\end{equation}
The chirp is weak and satisfy (\ref{slow-variation}), then
the equations for complex amplitudes are valid when
\begin{equation}\label{slow}
\alpha L\ll\kappa.
\end{equation}

Note that set (\ref{second-order}) is symmetric under
transformation $\alpha\to-\alpha, q\to-q,a\leftrightarrow
b$. Then the right reflection coefficient can be obtained
from the expression for left one by changing signs of
parameters $\alpha$ and $q$.

\section{Solution}\label{sect:3}

Equations (\ref{second-order}) are reduced to the confluent
hypergeometric form by the substitution
$t=\i\alpha(z-z_0)^2/2$:
\[
t\Ddot{a}+\left(\frac12- t\right)\Dot{a}+\i\eta
a=0,
\]
where dot denotes the derivative with respect to new
variable $t$, $\eta=\beta^2k_0^2/2\alpha$ is the adiabatic
parameter. The equation for second amplitude $b$ is complex
conjugated. The general solutions at $-L<z<L$ are linear
combinations
\begin{equation}\label{a-first}
    a(z)=A_1u_1(z)+A_2u_2(z),\quad
    b(z)=B_1u_1^*(z)+B_2u_2^*(z)
\end{equation}
of the Kummer confluent hypergeometric functions \cite{BE53}:
\begin{eqnarray}
u_1=F\left(-\i\eta;{\scriptstyle\frac12};\i\alpha(z-z_0)^2/2\right),\nonumber\\
u_2=(z-z_0) F\left({\scriptstyle\frac12}
-\i\eta;{\scriptstyle\frac32};\i\alpha(z-z_0)^2/2\right);
\label{u12}\\
F(a;c;x)=1+\frac{a}{c}\frac{x}{1!}
+\frac{a(a+1)}{c(c+1)}\frac{x^2}{2!}+\dots,\nonumber
\end{eqnarray}
where $A_1,A_2,B_1,B_2$ are constants and the asterisk
denotes the complex conjugation. The solution
was obtained in \cite{osaMHW75} for optical waveguide.
The solution for coupled-wave equations with identical
signs has been obtained in the context of nonadiabatic population inversion
in two-level system \cite{aplH75}.

The relations between constants can be obtained from set
(\ref{interaction}) near resonance point $z=z_0$ where
$a=A_1+A_2(z-z_0)+O(z-z_0)^2, b=B_1+B_2(z-z_0)+O(z-z_0)^2$:
\begin{equation}\label{constants}
\frac{A_2}{B_1}=\i k_0\beta\e^{\i\theta_0-\i\alpha
z_0^{2}/2},\quad
\frac{B_2}{A_1}=-\i
k_0\beta\e^{-\i\theta_0+\i\alpha z_0^{2}/2}.
\end{equation}
The right boundary condition (\ref{boundary}) yields the ratio of coefficients
$A_1$ and $A_2$
\begin{equation}\label{rho}
    \rho=\frac{A_1}{A_2}=\frac1{k_0^2\beta^2}\frac{B_2}{B_1}=-\frac
    {
      F\left(\i\eta;{\scriptstyle\frac12};-\i\alpha(L-z_0)^2/2\right)
    }
    {\beta^2k_0^2(L-z_0) F\left({\scriptstyle\frac12}
    +\i\eta;{\scriptstyle\frac32};-\i\alpha(L-z_0)^2/2\right)}.
\end{equation}
The left reflection and transmission coefficients
(\ref{transmission}) can be expressed in terms of confluent
hypergeometric functions
\begin{equation}\label{reflection-general}
    r=\frac{\e^{-\i\theta_0+\i\alpha z_0^{2}/2}}{\i k_0\beta}
    \frac{ u_1^*(-L)+\beta^2k_0^2\rho u_2^*(-L)}
    {\rho u_1(-L)+u_2(-L)},\quad
    t=\frac{\rho u_1(L)+u_2(L)}{\rho u_1(-L)+u_2(-L)},
\end{equation}
where functions $u_{1,2}$ are defined by (\ref{u12}) and $\rho$ is defined by (\ref{rho}).

\begin{figure}\centering
\includegraphics[width=0.75\textwidth]{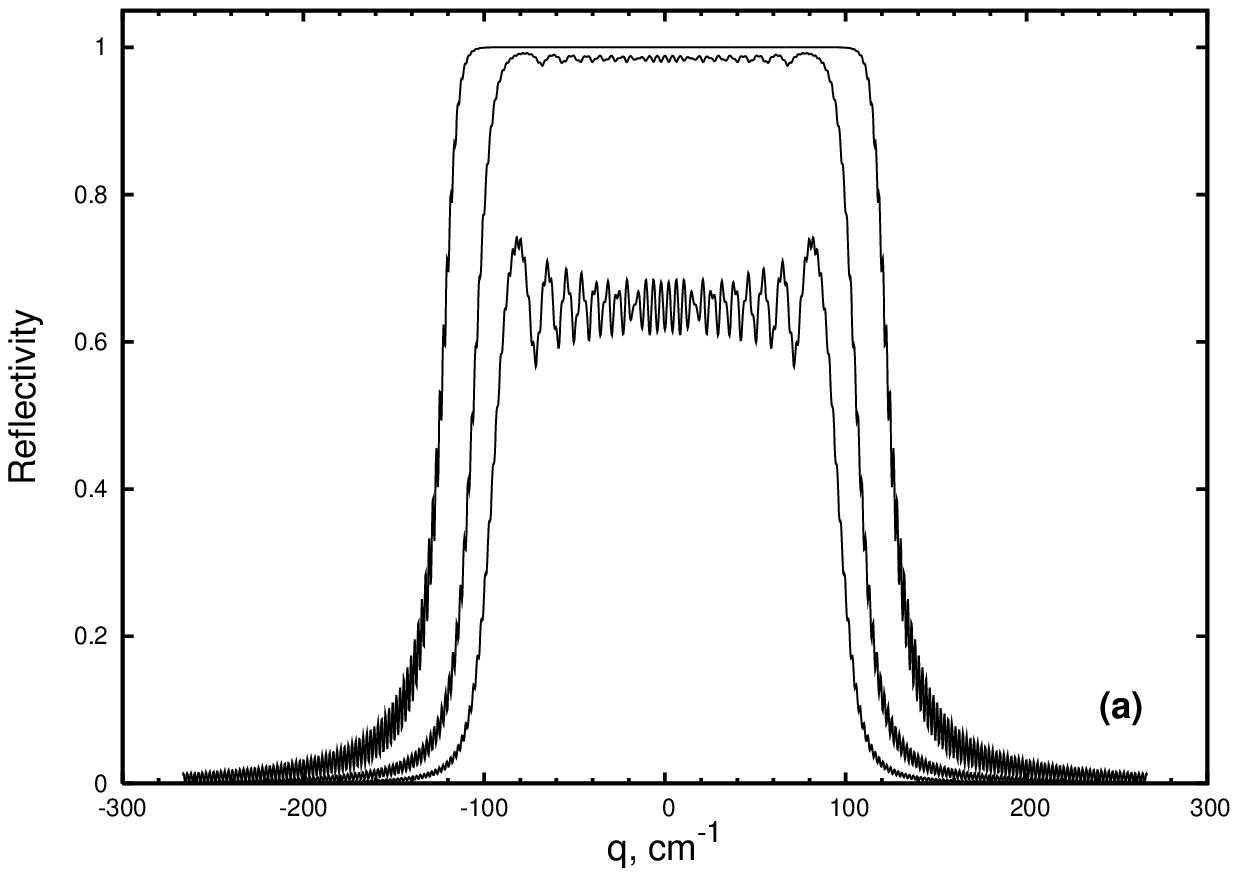}
\includegraphics[width=0.75\textwidth]{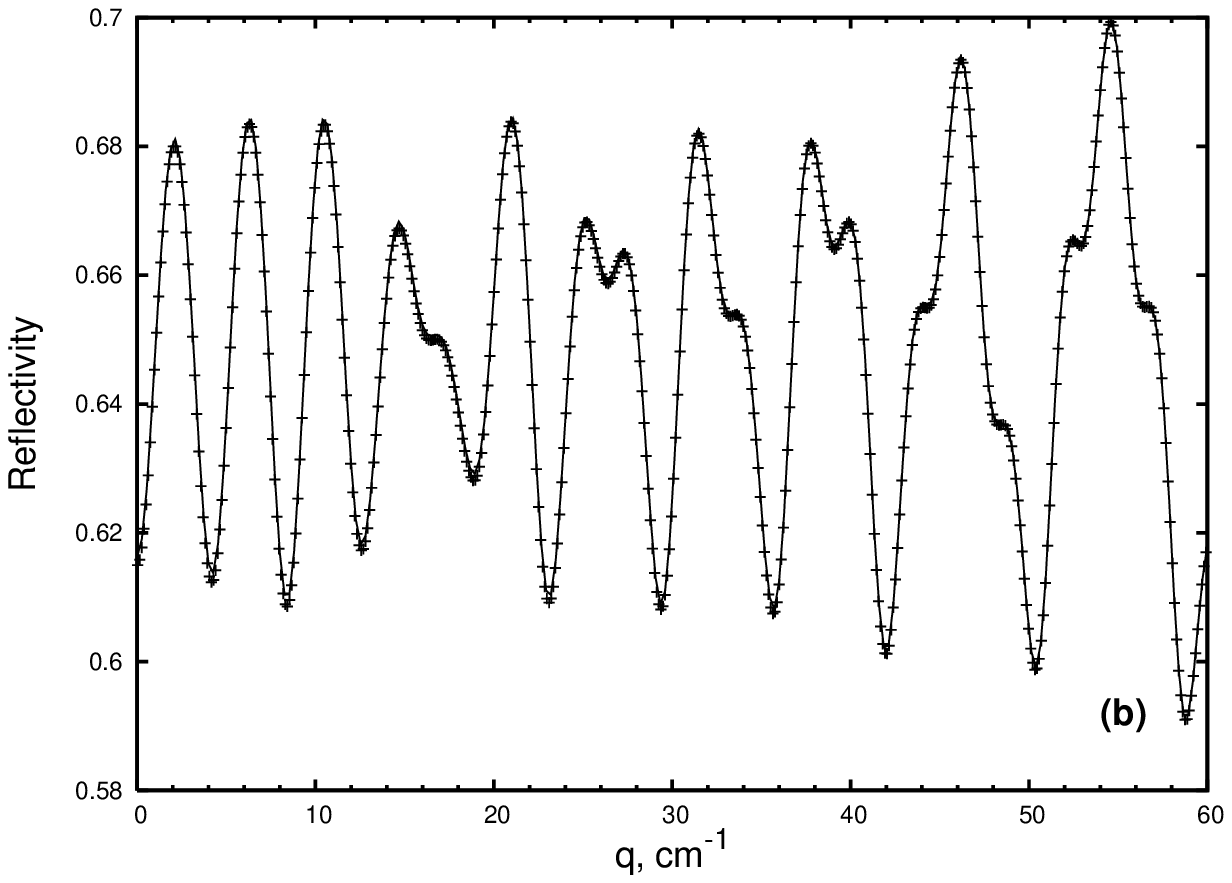}
\caption{(a) Reflection spectrum $R(q)$ at
$\alpha=600~\mbox{cm}^{-2}, L=0.5~\mbox{cm},
k_0=6\times10^{4}~\mbox{cm}^{-1}$, from the top down
$\beta=\beta_0, \beta_0/2, \beta_0/4$, where $\beta_0=0.67\times10^{-3}$. (b) A part of the lower curve
$\beta=\beta_0/4$, crosses denote the numerical
result. }\label{fig:spectrum}
\end{figure}

\begin{figure}\centering
\includegraphics[width=0.75\textwidth]{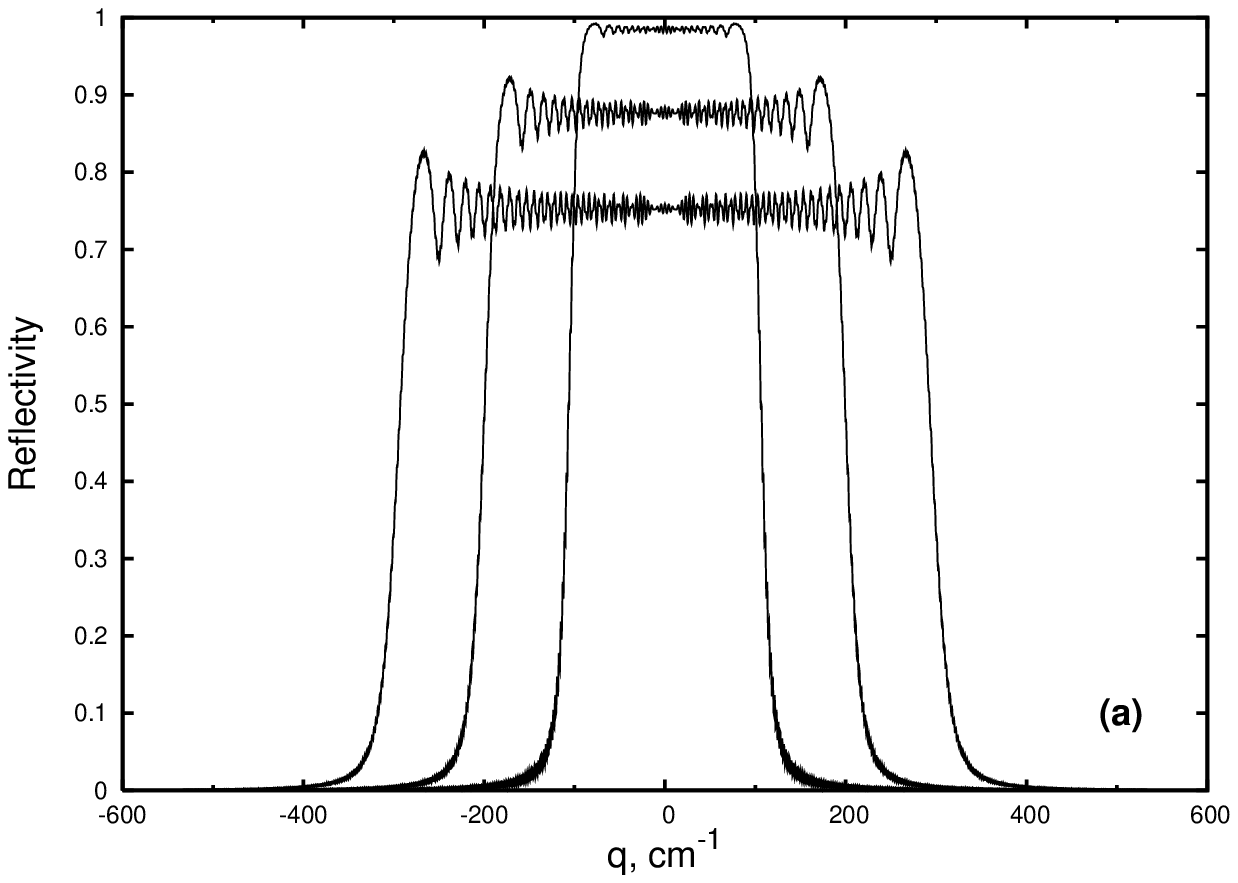}
\includegraphics[width=0.75\textwidth]{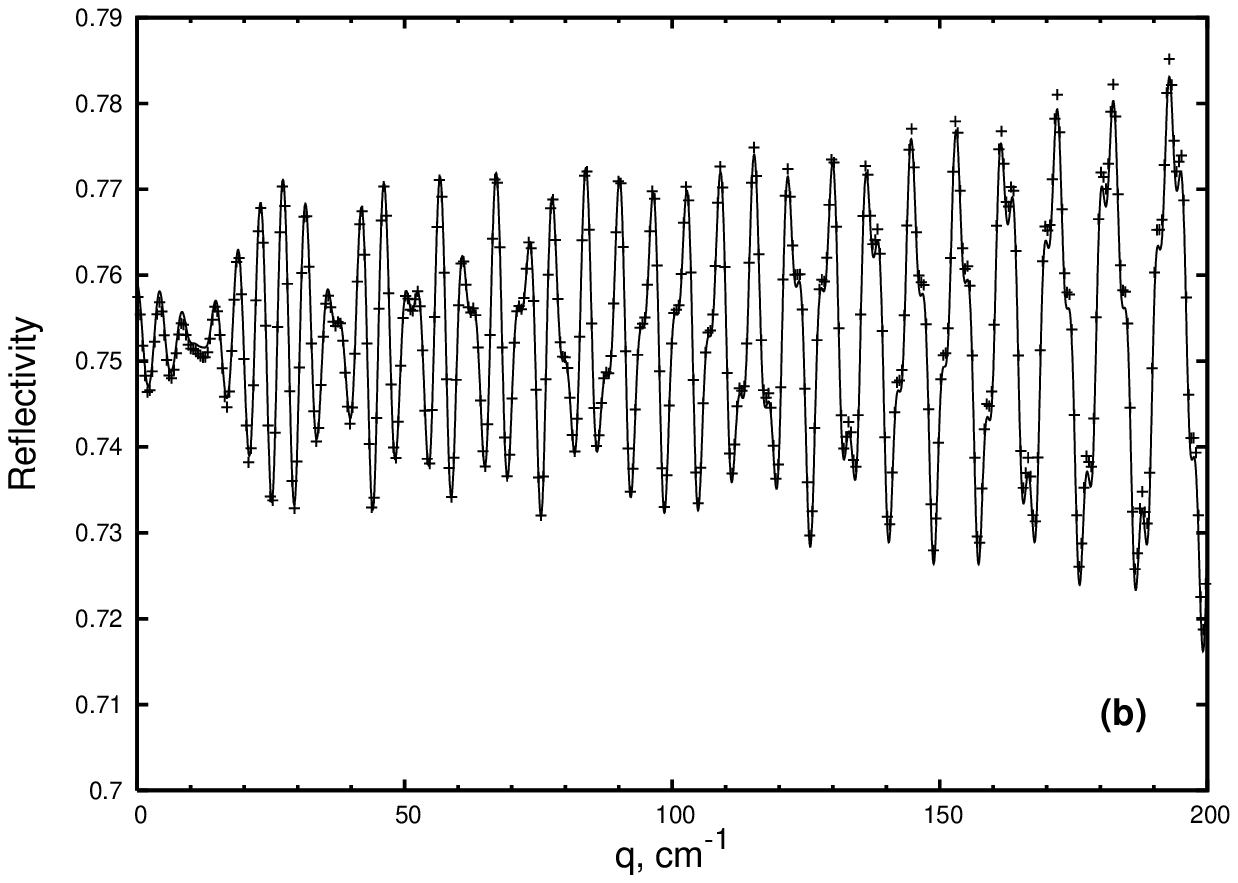}
\caption{(a) Reflection spectrum from the top down at
$L=0.5~\mbox{cm}, \beta=0.33\times10^{-3},
k_0=6\times10^{4}~\mbox{cm}^{-1}$ and $\alpha=\alpha_0,
2\alpha_0, 3\alpha_0,$,   where
$\alpha_0=600~\mbox{cm}^{-2}$. (b) A part of the lower
curve $\alpha=3\alpha_0$, crosses denote the numerical
result. }\label{fig:alpha-dependence}
\end{figure}

The reflection spectrum, i.e., the reflectivity $R=|r|^2$
as a function of detuning $q$, is shown in
Fig.~\ref{fig:spectrum} (a). The central frequency of the
spectrum comes to resonance at $z=0$, in the middle of
grating. The central part has a flat top at high adiabatic
parameter, as the upper curve shows, and the
maximal reflectivity is close to 1. The width of central
part is proportional to the length $L$. The reflectivity is
relatively high if the turning point $z_0$ lies inside the
grating $|z_0|<L$. This inequality gives the bandwidth
$|q|<\alpha L/2$. There is no parametric resonance at higher
detuning, when $|z_0|>L$, and the reflectivity is small.
Fig.~\ref{fig:alpha-dependence}~(a)
shows how the bandwidth grows up with the chirp parameter
$\alpha$ at fixed modulation depth $\beta$. The adiabatic
parameter decreases with $\alpha$, then the reflectivity in
the center decreases from curve to curve.

The spectrum was recalculated numerically by $T$-matrix
approach. The initial Helmholtz equation (\ref{Helmholtz})
was solved numerically with neither approximation of slow
envelope, nor quadratic term $(\delta n/n)^2$ neglecting.
The number of layers per period of spatial modulation was
fixed at $N=32$, then the step varied along the grating.
The spectra for $n=1.5$ and the same parameters are shown
in Fig.~\ref{fig:spectrum}~(b) by crosses. The numerical
results are very close to analytical, since both
dimensionless parameters controlling the validity of
coupled-wave approximation are small: $\beta\sim10^{-3},
\alpha L/\kappa\sim 5\times10^{-3}$. At higher parameter
$\alpha$ the deviation of coupled-wave equations solutions from
that of Helmholtz equation increases, but not dramatically, as
shown in Fig.~\ref{fig:alpha-dependence}~(b). The origin of the deviation
is resonance approximation (\ref{resonance-condition}). We replace $k$ by $k_0$ in coupled-mode equations (\ref{interaction}), while the Helmholtz wave equation (\ref{Helmholtz}) involves $k$. Comparing Fig.~\ref{fig:spectrum}~(a) and
Fig.~\ref{fig:alpha-dependence}~(b) we see that the latter involves higher detuning,
then the deviation is greater at higher $q=k-k_0$.

\begin{figure}\centering
\includegraphics[width=0.75\textwidth]{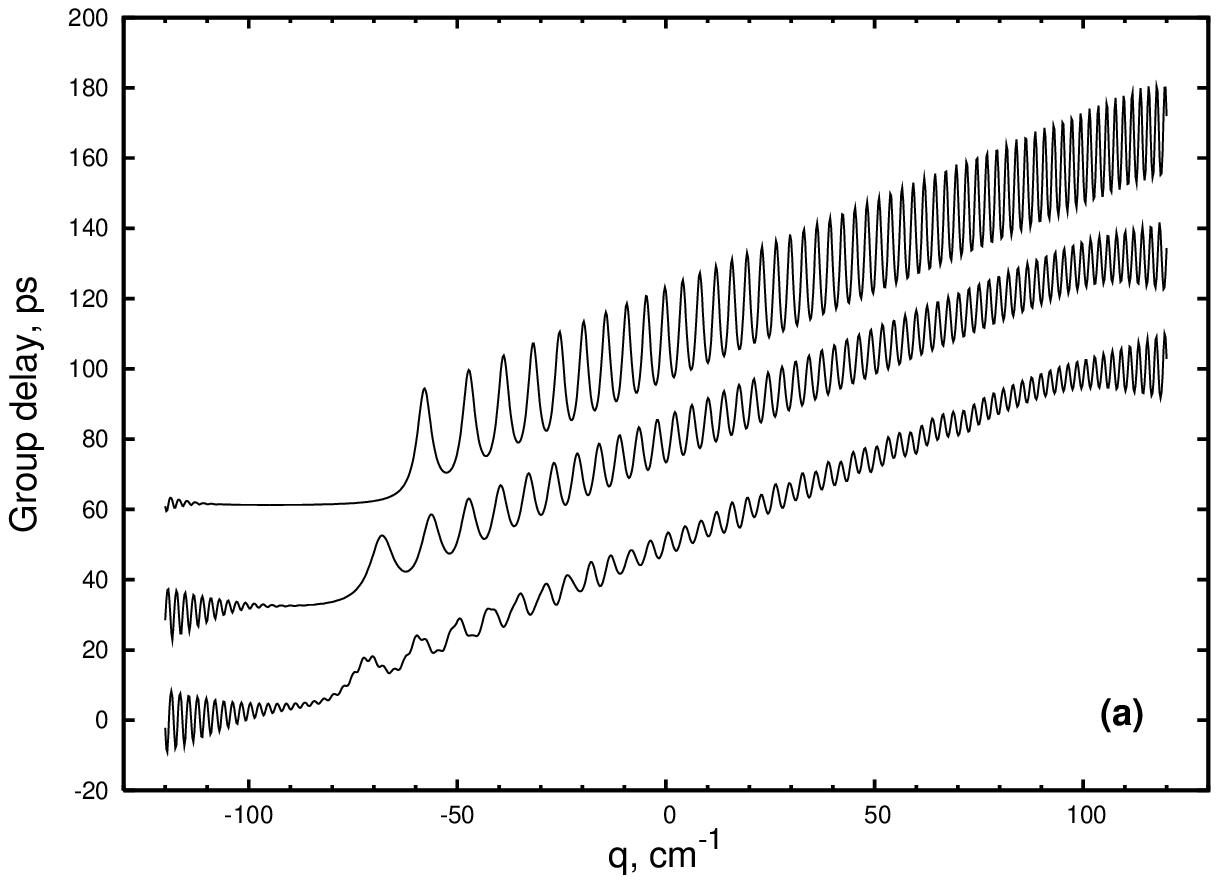}
\includegraphics[width=0.75\textwidth]{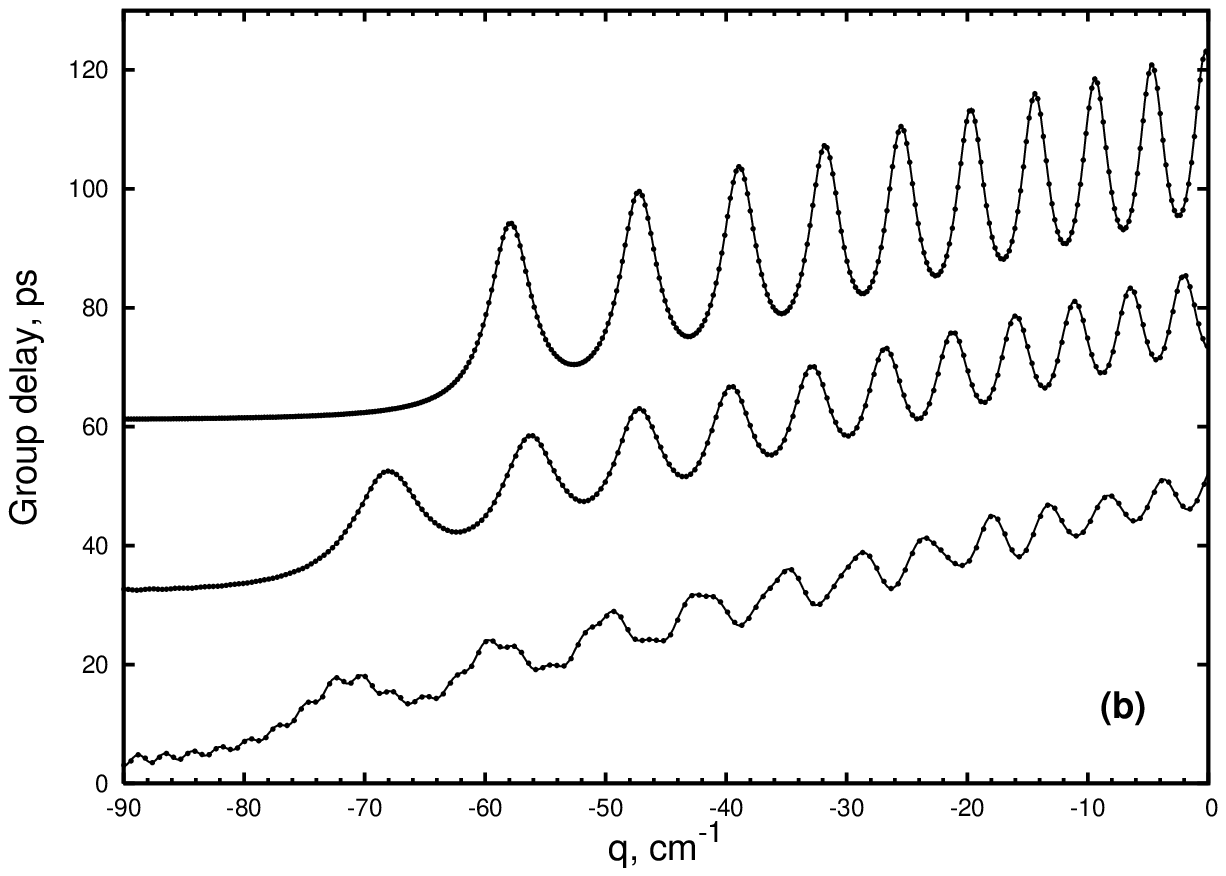}
\caption{(a) Group delay (ps) as a function of detuning $q$
(cm$^{-1}$) at $\alpha=600~\mbox{cm}^{-2}, L=0.5~\mbox{cm},
k_0=6\times10^{4}~\mbox{cm}^{-1}$, from the top down
$\beta=\beta_0, \beta_0/2, \beta_0/4$, where $\beta_0=0.67\times10^{-3}$. (b) Comparing with
numerical calculation denoted by
dots.}\label{fig:groupdelay}
\end{figure}

The group delay found from analytical solution
(\ref{reflection-general}) is
plotted in Fig.~\ref{fig:groupdelay}~(a) at the same
parameters as the reflection spectrum in
Fig.~\ref{fig:spectrum}. The deviation of curves from the
linear dependence, the group delay ripple, manifests itself
as oscillations with variable frequency. The frequency
grows up towards the blue end of spectrum in agreement with
results from \cite{ECOC97,RK99}. For the negative chirp (or when the incident light enters from the right) the frequency grows up towards the red edge of spectrum.
The maximum deviation from the averaged slope decreases with decreasing
modulation depth $\beta$. Meanwhile, the ripple in
reflectivity increases for small $\beta$. A fragment of
group delay characteristics is shown in
Fig.~\ref{fig:groupdelay}~(b) along with numerical
calculations. Dots obtained from numerical calculation
are very close to the curve given by
analytical formula.

It is difficult to analyze the solution in its general
form. In particular the cumbersome expression for group
delay, the derivative of (\ref{reflection-general}) with
respect to the detuning, is not presented here.  Let us
simplify expressions using the asymptotics of Kummer
functions in the next section.

\section{Asymptotics}\label{sect:4}

The asymptotic expressions for the reflection coefficient
can be obtained from (\ref{reflection-general}) in two
cases. The first case is the resonance condition at the
left end, namely, detuning $q=-\alpha L/2$ for which
$z_0=-L$. In this case it follows from (\ref{u12}) that
$u_1(-L)=1,u_2(-L)=0$, and then from
(\ref{reflection-general})
\begin{eqnarray}\label{tanh}
    r\approx\frac{\e^{-\i\theta_0+\i\alpha L^2/2}}{\i k\beta\rho}
    \approx-\frac{\e^{-\i\theta_0+\i\alpha L^2/2}}{\sqrt{\i\eta}}
    \frac{\Gamma(1/2-\i\eta)}{\Gamma(-\i\eta)},\\
       R=|r|^2\approx\tanh\pi\eta.\label{semi-infinite}
\end{eqnarray}

The other case is when the resonance point $z_0$ being far
from both ends inside the grating: $|q|<\alpha L/2$ and
$\alpha(z_0\pm L)^2/2\gg1$. The asymptotic expression of
the confluent hypergeometric functions \cite{BE53} at
$|\mathrm{arg}\,x|<\pi$
\begin{equation}\label{Kummer-asymptotics}
F(a;c;x)\approx\frac{\Gamma(c)}{\Gamma(c-a)}
\left(\frac{\e^{\i\pi\epsilon}}{x}\right)^a
+
\frac{\Gamma(c)}{\Gamma(a)}\e^xx^{a-c},\quad \epsilon=
\begin{cases}
+1, & \Im x>0,\\
-1, & \Im x<0
\end{cases}
\end{equation}
allows one to simplify expression (\ref{reflection-general}).

\begin{figure}\centering
\includegraphics[width=0.55\textwidth]{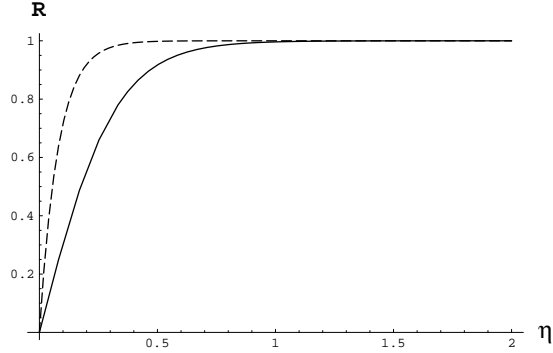}
\caption{The reflectivity $R$ as a function of adiabatic
parameter $\eta$ at $z_0=-L$
(solid line) and at $z_0=0$ (dashed).}\label{fig:saturation}
\end{figure}

The reflection coefficient can be written using (\ref{u12})
\begin{eqnarray}
    r\approx-\sqrt{R_0}\e^{\i\phi-\i\pi/4+2\i q^2/\alpha-\i\theta_0}
    \left[\frac\alpha2\left(L+
    \frac{2q}\alpha\right)^2\right]^{-2\i\eta}\times\nonumber\\
    \frac
    {
      1+\e^{+\i\pi/4-\i\phi}\sqrt{\frac{\eta}{R_0}}
\left(
      \psi_+^{2\i\eta-1/2}\e^{-\i\psi_+}+
      \psi_-^{2\i\eta-1/2}\e^{-\i\psi_-}
\right)
         }
    {
       1+\sqrt{\eta R_0}
       \left(
        \psi_+^{-2\i\eta-1/2}\e^{\i\psi_++\i\phi-\i\pi/4}+
\psi_-^{2\i\eta-1/2}\e^{-\i\psi_--\i\phi+\i\pi/4}
        \right)
    },
\label{reflection-simple}
\end{eqnarray}
where $\psi_\pm(q)=\alpha(L\pm2q/\alpha)^2/2$,
$\phi=\mbox{arg}[\Gamma(\i\eta)\Gamma(1/2+\i\eta)]$,
$R_0=1-\e^{-4\pi\eta}$ and we omit terms of the order of
$1/\alpha L^2\ll1$. The enumerator and denominator of the
fraction in the second line of (\ref{reflection-simple})
are close to 1, if $\psi_\pm\gg1$. Then the formula for
reflectivity becomes simple
\begin{equation}\label{4tanh}
    R\approx R_0=1-\e^{-4\pi\eta}.
\end{equation}
Both curves (\ref{semi-infinite}) and (\ref{4tanh}) are shown in
Fig.~\ref{fig:saturation}. One can see their saturation,
moreover, when the turning point $z_0=-L$, the saturation
occurs later than when the turning point is far from both
ends.

The group delay obtained from
(\ref{reflection-simple}) is
\begin{eqnarray}
    f=\frac{d\arg r}{d \omega}=\frac{n}{c}\frac{d\Im\ln r}{d q}
    \approx\frac{n}{c}
\Bigl\{\frac{4q}{\alpha}
    -2\sqrt{\frac{2\eta}{R_0\alpha}}\times\nonumber\\
    \times
\Bigl[
    (1+R_0)\cos\left(\phi-\pi/4+\psi_+-
    2\eta\ln\psi_+\right)+\nonumber\\
    +(1-R_0)\cos\left(\phi-\pi/4+\psi_--
    2\eta\ln\psi_-\right)
\Bigr] \Bigr\},\label{group_delay}
\end{eqnarray}
where we neglect terms of the order of $1/\alpha L^2\ll1$.
Expression (\ref{group_delay}) involves three terms. The
first (in the first line) gives the averaged slope. It is a linear function
within the bandwidth. Its slope depends on parameter
$\alpha$. At $1-R_0\ll1$ the second term (the second line) gives the ripple, chirped
oscillations.  The frequency of these oscillations is
double distance from left end of the grating to the
reflection point $z_0$. Their frequency
$\psi_+'(q)=2L+4q/\alpha$ grows up towards the blue edge of
the spectrum. When reflectivity $R_0$ becomes smaller, the
last term (the third line in Eq.\ref{group_delay}) proportional
to $T_0=1-R_0$ comes into effect. It
gives the additional oscillations with variable frequency
$\psi_-'(q)=2L-4q/\alpha$ that grows up towards the red
edge of the spectrum. It is precisely the sum of two chirped
oscillations with significantly different frequencies that
the left part of the lower curve in
Fig.~\ref{fig:groupdelay}~(a) displays. Magnified view of the corresponding
fragment is also shown in Fig.~\ref{fig:groupdelay}~(b).
If we change the sign of chirp parameter $\alpha$, then functions $\psi_\pm$
switch their roles: $\psi_+\leftrightarrow\psi_-$. Therefore at high reflectivity
$1-R_0\ll1$ the spatial frequency of leading oscillations $\psi_+'=2L+4q/\alpha$
decreases towards the shorter wavelengths.

\begin{figure}\centering
\includegraphics[width=0.75\textwidth]{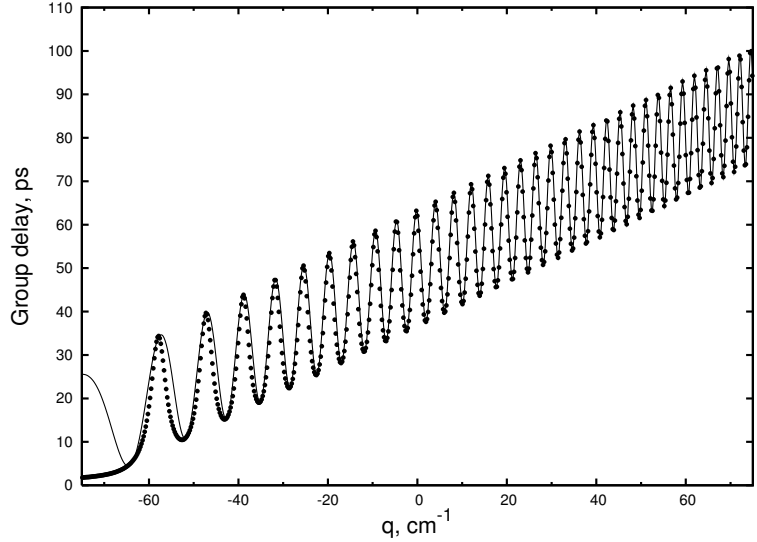}
\caption{The group delay calculated according to asymptotic
formula (\protect\ref{group_delay}) at
$\alpha=600~\mbox{cm}^{-2}, L=0.5~\mbox{cm},
k_0=6\times10^{4}~\mbox{cm}^{-1}$ $\beta=0.67\times10^{-3}$.
Dots denote the exact Kummer
solution.}\label{fig:oscillations}
\end{figure}

\begin{figure}\centering
\includegraphics[width=0.75\textwidth]{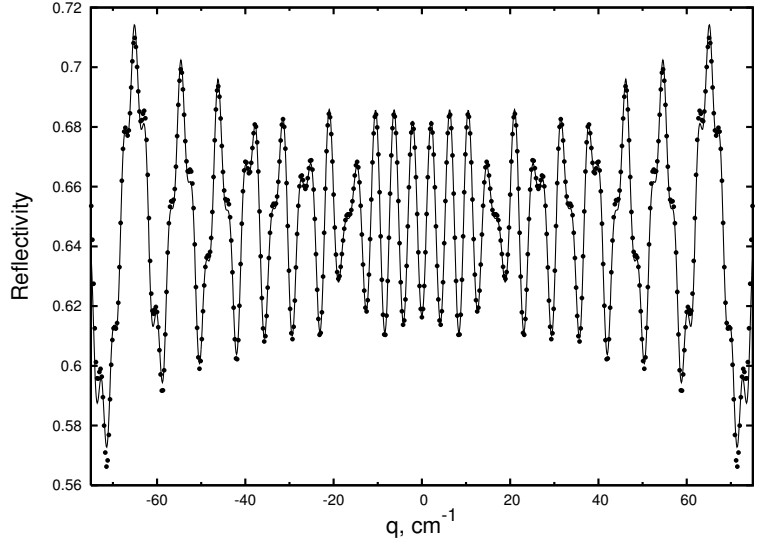}
\caption{Reflection spectrum calculated by asymptotic
formula (\protect\ref{modulation}) at $\alpha=600\mbox{
cm}^{-1}, \beta= 0.17\times10^{-3}, k_0 =
6\times10^{4}\mbox{ cm}^{-1}$. Dots denote the exact Kummer
solution.}\label{fig:asymptotics}
\end{figure}

The amplitude of oscillations in group delay
(\ref{group_delay}) increases when $R_0$ tends to unity,
while that in the spectrum decreases. Formula for the
reflection inside the bandwidth can be obtained from
(\ref{reflection-simple}) with the accuracy to the next
order of transparency $T_0=1-R_0$
\begin{eqnarray}
    R\approx R_0+2\sqrt{\eta R_0}(1-R_0)\times\nonumber\\
    \left[
    \frac{\cos(\phi-\pi/4+\psi_+-2\eta\ln\psi_+)}{\sqrt{\alpha/2}(L+2q/\alpha)}+
    \frac{\cos(\phi-\pi/4+\psi_--2\eta\ln\psi_-)}{\sqrt{\alpha/2}(L-2q/\alpha)}
    \right].
    \label{modulation}
\end{eqnarray}
At high reflectivity $R_0\to1$ oscillations
(\ref{modulation}) are suppressed. The first term in square
brackets describes oscillations with frequency
$2L+4q/\alpha$, their amplitude gains towards the red edge
of the spectrum. The second term corresponds to oscillations
with frequency $2L-4q/\alpha$ with amplitude
growing towards the blue edge. Both approximate formulas
(\ref{modulation}) and (\ref{group_delay}) for oscillations
are plotted in Fig.~\ref{fig:oscillations} and
Fig.~\ref{fig:asymptotics}, respectively. As figures
illustrate, the asymptotic expressions nearly coincide with
exact Kummer solutions.
The departure of the simple formula from the Kummer solution (left edge in Fig.~\ref{fig:oscillations} and both edges in Fig.~\ref{fig:asymptotics})
occurs when we get the limit of applicability of the asymptotic expansion. The turning point should be located far from the ends of grating, i.e., $|L\pm2q/\alpha|\gg L_{eff}=(2\pi/|\alpha|)^{1/2}$. Asymptotics are broken when the turning point occurs too close to the end.

The dependence on parameters $\alpha,\beta$ in
Fig.~\ref{fig:spectrum} and \ref{fig:alpha-dependence} can
also be explained by the asymptotic expressions. At fixed
chirp parameter $\alpha$ the adiabatic parameter
$\eta=(k_0\beta)^2/2\alpha$ in (\ref{4tanh}) decreases with
decreasing the modulation depth $\beta$. Then reflectivity
$R_0$ at $q=0$ is relatively small and oscillations
with amplitude $1-R_0$ in the spectrum become noticeable. At fixed
$\beta$, on the contrary, the adiabatic parameter decreases
with increasing $\alpha$. It is the reason of the most
evident oscillation in the spectrum corresponding to the
higher chirp parameter $\alpha$.

\section{Discussion}\label{sect:5}

The reflectivity is maximal
at $k=k_0=\kappa/2=\pi/\Lambda$, where $\Lambda$ is the period of
modulation in the middle of the grating, at $z=0$.
The spatial frequency of modulation $\theta'(z)=\alpha z+\kappa$ depends on
coordinate $z$. Then at some distance from the center the
wave with $k=k_0$ comes out from the resonance. The
dephasing occurs when $\theta=\alpha z^2/2\sim\pi$, i.e.,
at distance $z=L_{\textrm{eff}}\sim (2\pi/\alpha)^{1/2}$. The
effective number of strokes along length $L_{\textrm{eff}}$
should be large $M_{\textrm{eff}}=L_{\textrm{eff}}/\Lambda=(2\pi/\alpha)^{1/2}\Lambda^{-1}\gg1$.
Moreover, to provide the high reflectivity it
should satisfy the stricter limitation of dense grating
$M_{\textrm{eff}}\beta\gtrsim1$. From here we get a condition
for adiabatic parameter
\[
\eta=\frac{\beta^2k_0^2}{2\alpha}\gtrsim1/4\pi.
\]

The bandwidth of the reflection spectrum is $\alpha L$, as
shown in Sec.~\ref{sect:3}. The fronts of spectrum are
determined by the effective length $L_{\textrm{eff}}$. When
point $z_0=2q/\alpha$ is placed outside the grating at
distance $\sim L_{\textrm{eff}}$ from the end, the
reflection almost vanishes. The width of fronts is $\delta
q=\alpha L_{\textrm{eff}}=1/L_{\textrm{eff}}$. The fronts
are steep while $L_{\textrm{eff}}\ll L$, i.e., in the limit
of long grating.

Phase modulation $\theta(z)$ provides the parametric
resonance condition for different wavelengths. The shorter
waves meet their resonance at longer distance
$z_0=2q/\alpha$, and then the group delay of blue light is
more than that of red one, Fig.~\ref{fig:groupdelay}. The
linear dependence of the average group delay
(\ref{group_delay}) upon the detuning has also simple
explanation. The delay $\tau=f/v_{\textrm{gr}}$ is defined
by double distance from starting point to the resonance for
given wavenumber $f\approx2z_0=4(k-k_0)/\alpha$. Here
$v_{\textrm{gr}}$ is the group velocity of light. If the
chirp $\alpha$ is negative, then the sign of delay
characteristics becomes negative.

\begin{figure}\centering
\includegraphics[width=0.55\textwidth]{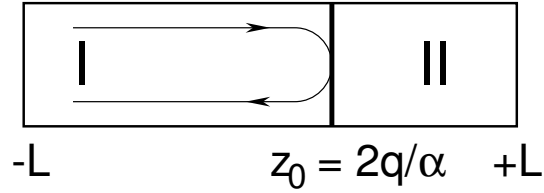}
\caption{Configuration of compound cavity: left ``mirror''
$z=-L$ is the left edge of grating, right ``mirror'' $z=+L$
is the right edge of grating. Middle variable ``mirror'',
the turning point $z=z_0$, is located at different
positions depending on the wavelength. Then the ripple
frequencies are determined by the variable lengths of
sub-cavities I and II.}\label{fig:compound}
\end{figure}

The ripple outside the reflection spectrum bandwidth,
Fig.~\ref{fig:spectrum},\ref{fig:alpha-dependence}, with
period $\pi/L$ are the Gibbs oscillations originated by
steep boundaries, i.e., reflection from the grating edges.
The aperiodic oscillation inside the bandwidth arise from
the triple-mirror cavity with moving middle mirror,
Fig.~\ref{fig:compound}. The wave reflected to the left
from turning point $z_0$ could reflect back to the right
from the left end of the grating. Then the cavity appears
between $z=-L$ and $z=z_0$; its effective length is
$l=z_0+L=2q/\alpha+L$. It results in oscillations with
period $\pi/(L+2q/\alpha)$. The cavity with variable
``mirror'' is longer for blue spectrum and shorter for red,
then the frequency of oscillations increases with $q$,
as mentioned in paper \cite{ECOC97}. At
$R_0\to1$ these oscillations are suppressed in the
reflection spectrum, but remain in the group delay
characteristics. If the reflectivity is not close to 1, the
additional oscillations come into effect due to the
``right'' cavity with variable ``left mirror''. Their
period $\pi/(L-2q/\alpha)$ on the contrary is longer for
red spectrum. These oscillations are suppressed at
$R_0\to1$ both in reflection spectrum and group delay
characteristics.

\section{Conclusions}\label{sect:6}

Thus, the analysis of the reflection spectrum and group
delay of linearly chirped grating becomes simple if the
turning point $z_0=2q/\alpha$ is far from both ends of the
grating compared to the effective length
$L_{\textrm{eff}}=(2\pi/\alpha)^{1/2}$. Formulas for
reflectivity demonstrate the irregular oscillations in the
reflection spectrum when
the adiabatic parameter is not large. The oscillations are
aperiodic and their amplitude slowly increases from the
center of spectrum. The nature of the oscillations is
reflection in compound cavity with a mobile middle
``mirror''. There are two terms in asymptotic expression.
The first has a period $\pi/(L+z_0)$ (round trip in the
left sub-cavity), the second --- $\pi/(L-z_0)$ (round trip
in the right sub-cavity).  The oscillations in group delay
characteristics have the same origin. The difference is
that the right sub-cavity takes a negligible part in
forming the oscillations of group delay characteristics at
$R_0\to1$.

The amplitude of oscillations is suppressed at high chirp
parameter $\alpha$ even at fixed reflectivity. The
conservation of high reflectivity with increasing $\alpha$
requires increasing parameter $\beta$. In order to suppress
both oscillations one must choose as high the modulation
depth as possible, but the limitation exists in fiber Bragg
grating manufacturing. The alternative method to diminish
the unwanted echo might be to provide the signal dephasing
by apodization, i.e., smoothing the grating profile \cite{RK99}.

\section{Acknowledgments}
Authors are grateful to S.A.~Babin for fruitful
discussions. The work is partially supported
by the CRDF grant RUP1-1505-NO-05 and the Government
support program of the leading research schools
(NSh-7214.2006.2).


\end{document}